\documentclass{jpsj-suppl}
\usepackage{txfonts} 

\def\vect#1{{\mbox{\boldmath $#1$}}}

\def\ketv#1{\mid \hspace{-0.5mm} #1 \rangle}
\def\brav#1{\langle #1 \hspace{-1mm} \mid}

\title{Repulsive aspects of pairing correlation\\in nuclear fusion reaction}

\author{Shuichiro \textsc{Ebata}$^{1}$ and Takashi \textsc{Nakatsukasa}$^{2,3}$}

\inst{$^{1}$Meme Media Laboratory, Hokkaido University, Sapporo 060-0813, Japan\\
$^{2}$Center for Computational Sciences, University of Tsukuba, Tsukuba 305-8577, Japan\\
$^{3}$RIKEN Nishina Center, Wako 351-0198, Japan\\
}

\email{ebata@nucl.sci.hokudai.ac.jp}

\recdate{July 15, 2013}

\abst{
Numerical simulation on nuclear collisions are performed
using the canonical-basis time-dependent 
Hartree-Fock-Bogoliubov theory (Cb-TDHFB)
in the three-dimensional coordinate space. 
Comparing results of the Cb-TDHFB and 
the conventional time-dependent Hartree-Fock (TDHF) calculations,
we study effects of the pairing correlation on
fusion reaction of $^{22}$O+$^{22}$O, 
$^{52}$Ca+$^{52}$Ca, and $^{22}$O+$^{52}$Ca,
using the Skyrme SkM$^*$ functional and a contact-type pairing energy functional.
Although current results are yet preliminary, 
they may suggest that the pairing correlation could hinder 
the fusion probability at energies in the vicinity of the Coulomb barrier height. 
We also perform a calculation for heavier nuclei, $^{96}$Zn+$^{124}$Sn, 
which seems to suggest a similar hindrance effect. 
}

\kword{Heavy-ion collision, Pairing correlation,
Time-dependent mean-field theory}

\begin{document}
\maketitle
\section{Introduction}
The time-dependent Hartree-Fock theory (TDHF) is well-known as a useful tool to study nuclear dynamics. 
The studies of heavy-ion collision by using TDHF have been intensely performed since 1970's \cite{Neg82}. 
Recently, the TDHF calculation for the collision reaction with a realistic effective interaction 
in the three-dimensional coordinate space representation has become feasible 
with the help of the progress in computational power.
However, the TDHF cannot describe effects of pairing correlation
which plays an important role in nuclear structure and low-energy excitations. 
The time-dependent Hartree-Fock-Bogoliubov theory (TDHFB) 
is able to treat the static and dynamical pairing correlation self-consistently.
However, so far, no study of heavy-ion collision has been done using the TDHFB with a modern effective 
interaction in the three-dimensional space,
because of a number of numerical difficulties and requirement of the huge computational resources.
It should be noted that there have been a number of recent efforts toward
this direction \cite{BL10, Ha13}. 

In order to study nuclear dynamics treating the pairing correlation, 
we proposed the canonical-basis TDHFB (Cb-TDHFB)\cite{EN10}. 
The Cb-TDHFB is derived from full TDHFB equations represented in the canonical basis 
which diagonalize the density matrix, and using a BCS-like approximation for the pairing functional. 
We confirmed the validity of Cb-TDHFB for the linear response calculations, 
comparing the results with those of the quasi-particle random phase approximation 
which is a small amplitude limit of the full TDHFB\cite{EN10}.

In this study, we apply the Cb-TDHFB to the heavy-ion collision for symmetric cases ($^{22}$O+$^{22}$O, $^{52}$Ca+$^{52}$Ca) 
and for an asymmetric case ($^{22}$O+$^{52}$Ca).
We also make a preliminary study for a heavier case ($^{96}$Zn+$^{124}$Sn). 
The numerical simulation is performed in the three-dimensional Cartesian
coordinate space using the Skyrme effective interactions with the contact pairing functional. 
In Sec.2, we introduce the Cb-TDHFB equations,
and show the adopted pairing energy functional.
Then, in Sec.3, we discuss effects of the pairing correlation on the fusion reaction,
by comparing results of the Cb-TDHFB with those of the TDHF.

\section{Cb-TDHFB method}
\subsection{Cb-TDHFB equations and pairing functional}
The Cb-TDHFB equations can be derived from the TDHFB equations 
with an approximation for pairing functional\cite{EN10}. 
According to the Block-Messiah theorem\cite{RS80},
the TDHFB state at any time can be expressed in the canonical (BCS) form, 
\begin{eqnarray}
|\Phi (t) \rangle \equiv 
\prod_{l>0} \Big( u_{l}(t) + v_{l}(t) \hat{c}_{l}^{\dag}(t)\hat{c}_{\bar l}^{\dag}(t)  \Big) | 0 \rangle, 
\label{eq:BCS}
\end{eqnarray}
where $u_{l}(t),v_{l}(t)$ are time-dependent BCS factors and 
$\{ \hat{c}_{l}^{\dag}, \hat{c}_{\bar l}^{\dag}\}$ are creation operators of canonical pair of states $(l,\bar{l})$. 
In general, the time-evolution of the canonical states are
given by rather complex equations.
However, when the pair potential is diagonal, the equations are given in
a simple form.
Thus, we only take into account the ``diagonal'' parts of the pair potential,
\begin{eqnarray}
\Delta_{l}(t) = - \sum_{k>0} \kappa_{k}(t)\ \bar{\cal V}_{l\bar{l},k\bar{k}}(t) \ ,
\label{eq:delta}
\end{eqnarray}
where the pair probability $\kappa_{k}(t) \equiv u_{k}(t)v_{k}(t)$ 
corresponds to the pair tensor $\kappa (t)$ in the canonical-basis,
and $\bar{\cal V}_{l\bar{l},k\bar{k}}$ are the anti-symmetric two-body matrix elements.
Note that, since the canonical basis themselves evolve in time,
the two-body matrix elements $\bar{\cal V}_{l\bar{l},k\bar{k}}$
depend on time as well.
This leads to the Cb-TDHFB equations with the pair potential of
Eq.(\ref{eq:delta}), as follows. 
\begin{eqnarray}
i \hbar \frac{\partial \phi_{l}(t)}{\partial t} &=& \big\{ \hat{h}(t) - \eta_{l}(t) \big\} \ \phi_{l}(t),\ \ \ \ i \hbar \frac{\partial \phi_{\bar l}(t)}{\partial t}  = \big\{ \hat{h}(t) - \eta_{\bar{l}}(t)\big\} \ \phi_{\bar{l}}(t),  \nonumber \\
i \hbar \frac{\partial \rho_{l}(t)}{\partial t}  &=& \kappa_{l}(t)\Delta_{l}^{\ast}(t) - \kappa_{l}^{\ast}(t)\Delta_{l}(t),  \nonumber \\
i \hbar \frac{\partial \kappa_{l}(t)}{\partial t}  &=& \big\{ \eta_{l}(t) + \eta_{\bar l}(t) \big\}\ \kappa_{l}(t) + \Delta_{l}(t) \big\{2\rho_{l}(t) - 1 \big\}, 
\label{eq:Cb-TDHFB}
\end{eqnarray} 
where $\eta_{l}(t) \equiv \brav{\phi_{l}(t)} \hat{h}(t)\! \ketv{\phi_{l}(t)} + i \hbar \brav{\frac{\partial \phi_{l}}{\partial t}}\! \phi_{l}(t) \rangle$. 
The Cb-TDHFB equations determine the time evolution of
the canonical basis $\phi_{l}(t), \phi_{\bar l}(t)$, 
the occupation probability $\rho_{l}(t) \equiv |v_{l}(t)|^{2}$,
and the pair probability $\kappa_{l}(t)$. 
The equations conserve the orthonormal property of the canonical basis and the average particle number. 
When we choose a special gauge condition $\eta_{l}(t) = \varepsilon_{l}(t) = \brav{\phi_{l}(t)}\hat{h}(t) \! \ketv{\phi_{l}(t)}$, 
they also conserve the average total energy. 
At the static limit ($\partial\phi_l/\partial t = \partial \rho_l/\partial t
= \partial \kappa_l/\partial t = 0$),
they lead to HF+BCS ground state.
A boosted HF+BCS state is used as
the initial state ($t=0$) of the time evolution. 

We introduce neutron-neutron and proton-proton BCS pairing of a zero-range
contact type. 
The BCS pairing matrix elements ${\cal V}_{l\bar{l},k\bar{k}}^{\tau}$
are written as 
\begin{eqnarray}
 {\cal V}_{l\bar{l},k\bar{k}}^{\tau}=\int\! d\vect{r}_{1}d\vect{r}_{2} \sum_{\sigma_{1},\sigma_{2}} 
 \phi_{l}^{\ast}(\vect{r}_{1},\sigma_{1})\phi_{\bar{l}}^{\ast}(\vect{r}_{2},\sigma_{2})
 \hat{\cal V}^{\tau}(\vect{r}_{1},\sigma_{1};\vect{r}_{2},\sigma_{2}) \nonumber \\
 \times \left[  \phi_{k}(\vect{r}_{1},\sigma_{1})\phi_{\bar{k}}(\vect{r}_{2},\sigma_{2})
 -\phi_{\bar{k}}(\vect{r}_{1},\sigma_{1})\phi_{k}(\vect{r}_{2},\sigma_{2}) \right]. \hspace{-15mm}
\label{eq:Viijj0}
\end{eqnarray}
We introduce the spin-singlet contact interaction to Eq.(\ref{eq:Viijj0}): 
\begin{eqnarray}
\hat{\cal V}^{\tau}(\vect{r}_{1},\sigma_{1};\vect{r}_{2},\sigma_{2}) 
\equiv V_p^{\tau} \frac{1-\hat{\vect{\sigma}}_{1}\cdot \hat{\vect{\sigma}}_{2}}{4} \delta(\vect{r}_{1}-\vect{r}_{2}), 
\label{eq:Vint}
\end{eqnarray}
where the superscript $\tau$ distinguishes neutron and proton channels,
and $V_p^{\tau}$ is a strength of pairing functional \cite{KB90}. 
Here, we choose the simplest contact pairing functional (volume type)
for simplicity.

\subsection{Numerical details}
We calculate symmetric collisions, $^{22}$O+$^{22}$O and $^{52}$Ca+$^{52}$Ca, 
and an asymmetric collision $^{22}$O+ $^{52}$Ca,
using both the TDHF and Cb-TDHFB methods.
We adopt the Skyrme energy density functional of the SkM$^*$ parameter set.
We first prepare the ground states of projectile and target nuclei.
They are obtained by performing the self-consistent HF and HF+BCS calculation.
The center-of-mass correction is neglected in the present calculation. 
Then, the initial state of the simulation is constructed by locating
these two wave functions (projectile and target)
at a given impact parameter $b$ and at a relative distance $H$.
The distance $H$ should be large enough that
they interact through only the Coulomb interaction. 
We boost the wave functions with a given center-of-mass energy $E_\textrm{cm}$,
and calculate the time evolution according to Eq.(\ref{eq:Cb-TDHFB}). 

We use the three-dimensional Cartesian coordinate-space representation for the 
canonical states, $\phi_{l}(\vect{r},\sigma; t) = \brav {\vect{r},\sigma}\! \phi_{l}(t) \rangle$ 
with $\sigma=\pm 1/2$. 
The ground-state wave functions are obtained in the cubic box of (20 fm)$^3$.
These nuclei are selected because they have a spherical shape at the
ground state both in HF and HF+BCS calculations.
The space for the TDHF and Cb-TDHFB calculations 
is a rectangular box of 32 fm $\times$ 20 fm $\times$ 40 fm, 
discretized in the square mesh of $\Delta x = \Delta y = \Delta z = 1.0$ fm. 

In order to find the Coulomb barrier height,
we also perform the calculation
with the frozen density approximation at various distances between
the projectile and the target.
The calculated values of the Coulomb barrier height are
9 MeV, 49 MeV, and 21 MeV for
$^{22}$O+$^{22}$O and $^{52}$Ca+$^{52}$Ca, 
and $^{22}$O+$^{52}$Ca, respectively.
In this study, we choose the center-of-mass collision energy $E_\textrm{cm}$ 
near but slightly higher than these values.

\section{Results} 

\subsection{Symmetric collisions $\rm :\ ^{22}O$+$\rm ^{22}O$, $\rm ^{52}Ca$+$\rm ^{52}Ca$}
We simulate the $^{22}$O+$^{22}$O collision 
with an incident energy $E_{\rm cm}=$10 MeV.
The initial distance $H$ between projectile and target is 20 fm
along z-axis.
The impact parameter is varied as $b=$ 2.7$\sim$3.1 fm
in the $x$-axis direction. 
In $^{22}$O, the neutrons are in the superfluid phase, 
while the protons are in the normal phase.
The neutron pairing strength $V_{0}^{\rm n}$ is defined to reproduce
the experimental gap energy obtained from the binding energies
using the three-points formula.
The average neutron gap energy 
$\bar{\Delta}^{\rm n} \equiv \sum_{l>0} \Delta_{l}^{\rm n} / \sum_{l>0}$ 
is 2.06 MeV.

The results for $^{22}$O$+^{22}$O have been partially reported previously
\cite{EN14}.
In Figs. 1 and 2 in Ref. \cite{EN14},
the time evolution of neutron density distributions is presented 
for the $^{22}$O+$^{22}$O collision with a impact parameter $b$=3 fm 
in the simulations using TDHF and Cb-TDHFB. 
A remarkable difference is observed 
between the results with and without pairing correlation. 
In the TDHF calculation, a neck-like structure is formed leading to the fusion. 
In contrast, the neck formation does not take place in the Cb-TDHFB, 
and they do not fuse. 

In the case of Cb-TDHFB calculation, 
the results may depend on the initial choice of the gauge angle.
Namely, there is an additional degree of freedom for the phase of
the pair-probability $\kappa_l(t)$.
To investigate this, 
we change the phase of $v_l(t=0)$ of projectile (or target)
by an angle $\theta$, as $v_l(0) \rightarrow v_l(0)e^{i \theta}$.
Although we have not fully performed the investigation yet,
for the cases that the relative phase between projectile and target 
is $\theta=0$ and $\pi/4$, 
we confirm that the results are almost invariant. 

The fusion cross section $\sigma_{F}$ can be evaluated using a
semi-classical formula \cite{FL96}:
\begin{eqnarray}
\sigma_F (E) = \frac{\pi}{k^2} \sum_{L=0}^{L_{\rm max}} (2L+1) = \frac{\pi}{k^2}(L_{\rm max}+1)^2
\label{eq:sig}
\end{eqnarray}
where $L_{\rm max}$ is the maximum angular momentum for the fusion, 
and is evaluated as $L_{\rm max} \equiv k b_{\rm max}$ with 
the maximum impact parameter $b_{\rm max}$ and relative momentum $k$. 
The results for $^{22}$O+$^{22}$O indicate that
the fusion cross section calculated in TDHF (Cb-TDHFB) is larger (less)
than $\sigma_F$=11.79 $\pi$ fm$^2$.
To investigate further the pairing effect, we test a slightly weakened
strength of the pairing energy functional.
We have found that the fusion cross section 
in the weak pairing strength is lager than $\sigma_F$ with the original strength. 
These seem to indicate that the pairing correlation 
has a ``repulsive'' effect in the fusion reaction. 

\begin{figure}[h]
\begin{center}
\includegraphics[width=140mm, clip]{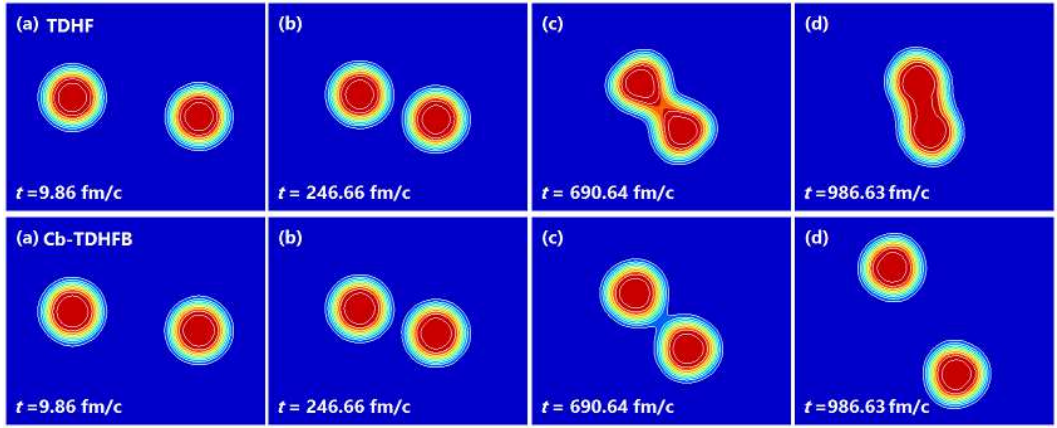}
\caption{(color on-line)
Neutron density distributions of $xz$-plane in $^{52}$Ca+$^{52}$Ca collision 
at $t$=($a$)9.86, ($b$)246.66, ($c$)690.64 and ($d$)986.63 ${\rm fm / c }$. 
Upper and lower panels indicate results of TDHF and Cb-TDHFB simulation 
with $E_{\rm cm}$=51.5 MeV and $b$=2.45 fm, respectively.}
\label{fig: 52Ca+52Ca}
\end{center}
\end{figure}

Next, let us discuss the heavier symmetric case of $^{52}$Ca+$^{52}$Ca.
Again, only the neutrons are in the superfluid phase with
an average gap energy of $\bar{\Delta}^{\rm n} = 1.86$ MeV.
We choose the collision energy $E_{\rm cm}=51.5$ MeV,
and vary the impact parameter as $b=$2.2$\sim$2.6 fm. 
In Fig.~\ref{fig: 52Ca+52Ca},
we show the time evolution of the neutron density distribution
in the $xz$-plane.
Before two nuclei touch (panels (a) and (b)), the TDHF and Cb-TDHFB simulations show almost same behavior. 
After the touching point, in Fig.~\ref{fig: 52Ca+52Ca} (c),
the difference appears in the neutron-density distribution. 
As is indicated in Fig.~\ref{fig: 52Ca+52Ca} (d),
two nuclei fuse in the TDHF while they do not in the Cb-TDHFB calculation. 
This is consistent with the case of $^{22}$O$+^{22}$O, 
though the difference in $\sigma_F$ between TDHF and Cb-TDHFB 
is smaller than the case of $^{22}$O.
In particular, the difference at the $^{22}$O simulation is 1.33 $\pi$ fm$^2$ (TDHF : Cb-TDHFB = 11.79 : 10.46), 
and for the $^{52}$Ca it is 0.28 $\pi$ fm$^2$ (TDHF : Cb-TDHFB = 6.78 : 6.50).

\subsection{Asymmetric collision $\rm :\ ^{22}O$+$\rm ^{52}Ca$}

\begin{figure}[h]
\begin{center}
\includegraphics[width=105mm, clip]{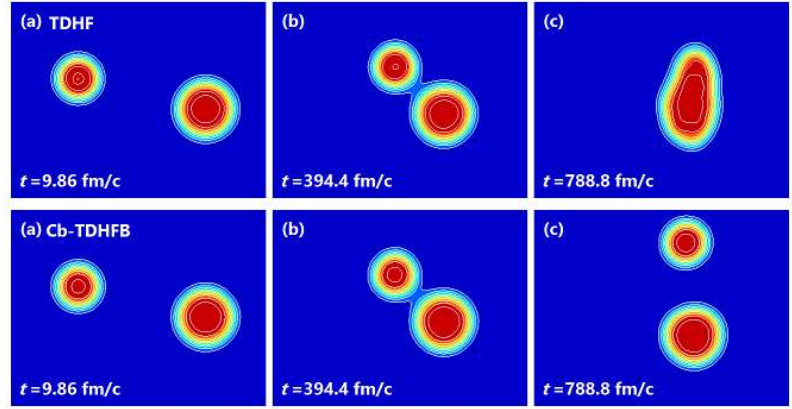}
\caption{(color on-line) Same as Fig. \ref{fig: 52Ca+52Ca}, but for $^{22}$O+$^{52}$Ca collision 
at $t$=($a$)9.86, ($b$)394.4 and ($c$)788.8 ${\rm fm / c }$. 
$E_{\rm cm}$=25 MeV and $b$=4.1 fm.}
\label{fig: 22O+52Ca}
\end{center}
\end{figure}

Next, we study the $^{22}$O+$^{52}$Ca collision.
The $E_{\rm cm}$ is taken as 25 MeV in this case.
The range of the impact parameter is from 3.0 to 4.5 fm. 
The pairing strength is chosen to be
the average value between $^{22}$O and $^{52}$Ca,
$\bar{V}_p^{\rm n}(^{22}{\rm O} + ^{52}{\rm Ca}) \equiv (V_p^{\rm n}(^{22}{\rm O})+V_p^{\rm n}(^{52}{\rm Ca}))/2$.
This produces the average gap energy larger (smaller) than the
experimental value for $^{22}$O ($^{52}$Ca).
Figure \ref{fig: 22O+52Ca} shows the neutron density distributions
same as Fig. \ref{fig: 52Ca+52Ca} but for $^{22}$O+$^{52}$Ca. 
We can see again a repulsive effect of pairing correlation
in the fusion reaction. 

\subsection{Collision of heavy nuclei $\rm :\ ^{96} Zr$+$\rm ^{124} Sn$}
It is known that the fusion reaction is hindered for heavier systems,
when the charge product of projectile and target is larger than about 1600,
$Z_{\rm P}Z_{\rm T}>1600$.
Here, we show our preliminary results for the head-on collision ($b=0$) of
$^{96}$Zr+$^{124}$Sn ($Z_{\rm P}Z_{\rm T}$=2000). 
In this preliminary study, the pairing strength is not yet well tuned
so that the calculated average neutron gaps,
$\bar{\Delta}^{\rm n} =$ 2.73 MeV for $^{96}$Zr and
2.42 MeV for $^{124}$Sn,
are about twice larger than the experimental values.
The SLy4d energy functional is adopted in this calculation.
For $^{124}$Sn, the calculated HF ground state no longer has a spherical
shape, but has an oblate shape ($\beta \sim -0.1$). 
Here, we choose the orientation that the symmetry axis of $^{124}$Sn is
perpendicular to the collision direction.
This produce the barrier height almost identical to that in the Cb-TDHFB
calculation with spherical $^{124}$Sn.
We only calculate the head-on collision ($b=0$)
in the rectangular box of 20 fm $\times$ 20 fm $\times$ 50 fm. 

\begin{figure}[h]
\begin{center}
\includegraphics[width=150mm, clip]{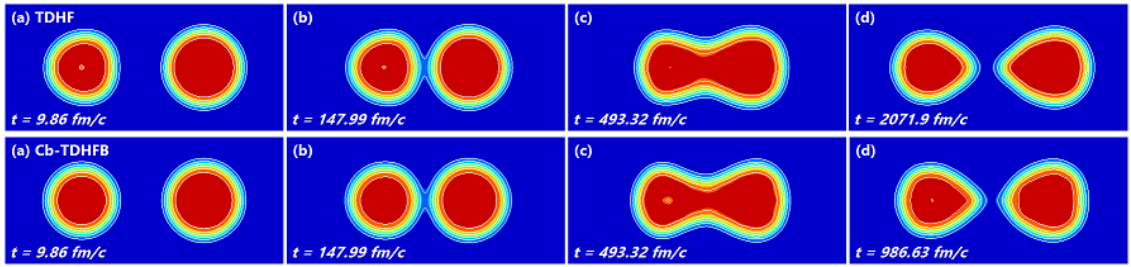}
\caption{(color on-line) Same as Fig. \ref{fig: 52Ca+52Ca},
but for $^{96}$Zr+$^{124}$Sn 
head on collision with $E_{\rm cm}$=227.7 MeV.
See the text for details.}
\label{fig: 96Zr+124Sn}
\end{center}
\end{figure}

Figure \ref{fig: 96Zr+124Sn} 
shows the snap shots of neutron density distribution for this case.
The panels (a-c) show the neutron density at $t\approx 10$ fm/c,
$t\approx 150$ fm/c (touching), and $t\approx 500$ fm/c (neck formation).
In this simulation, the system does not fuse.
The panel (d) shows the density profile for the scission point.
Although the obtained shapes in these panels are rather similar between
TDHF and Cb-TDHFB calculations,
we find a remarkable difference of the time duration from (b) to (d).
In the Cb-TDHFB calculation, the two nuclei departs again at
$t\approx 1000$ fm/c, while it happens at $t\approx 2000$ fm/c in the TDHF.
Although this result is still in a preliminary stage,
our results indicate that the impact of pairing correlation
exists in the heavier systems with the fusion hindrance
($Z_{\rm P}Z_{\rm T}>1600$).

\section{Summary}
We performed numerical simulations for the nuclear collisions of
symmetric combinations ($^{22}$O+ $^{22}$O, $^{52}$Ca+$^{52}$Ca), 
and the asymmetric one ($^{22}$O+$^{52}$Ca),
using the Cb-TDHFB theory including nuclear pairing correlation, 
in the three-dimensional Cartesian coordinate space. 
Comparing the results with those of the TDHF,
we discussed the impact of pairing correlations
in the low-energy nuclear reaction in the  vicinity of the Coulomb barrier. 

In these cases,
we have found that the fusion cross section obtained in Cb-TDHFB simulation is 
smaller than those of TDHF. 
These results may indicate the pairing correlation may hinder
the nuclear fusion probability.
However, in the Cb-TDHFB calculations, since we should check the initial gauge
angle dependence in more details,
our conclusion is still preliminary yet.

A calculation for $^{96}$Zr+$^{124}$Sn also shows an interesting difference
between TDHF and Cb-TDHFB.
In both cases, the two nuclei do not fuse.
However, in the Cb-TDHFB calculation, the two nuclei stay together
for much shorter time period.
This should be investigated further in future.

\section*{Acknowledgement} 
This work is supported by KAKENHI No. 24105006 and 25287065.
The numerical calculation has been performed at the high performance computing system 
at Research Center for Nuclear Physics, Osaka University.

\end{document}